\documentclass[aps,amsmath,amssymb,prl,twocolumn,showpacs,superscriptaddress]{revtex4-1}
\usepackage{bm}
\usepackage{epsfig}
\usepackage[usenames]{color}
\usepackage{graphicx}

\DeclareMathOperator{\Imm}{Im}

\DeclareMathOperator{\D}{\mathcal D}

\renewcommand{\paragraph}[1]{\textit{#1.---} }

\begin{document}

\title{{\it \bf Giant} Quantum Freezing of Tunnel Junctions mediated by Environments}

\author{A.~Glatz}
\affiliation{Materials Science Division, Argonne National Laboratory, Argonne, Illinois 60439, USA}

\author{N.~M.~Chtchelkachev}
\affiliation{Materials Science Division, Argonne National Laboratory, Argonne, Illinois 60439, USA}
\affiliation{Institute for High Pressure Physics, Russian Academy of Science, Troitsk 142190, Russia}
\affiliation{Department of Theoretical Physics, Moscow Institute of Physics and Technology, 141700 Moscow, Russia}

\author{I.~S.~Beloborodov}
\affiliation{Department of Physics and Astronomy, California State University Northridge, Northridge, CA 91330, USA}

\author{V.~Vinokur}
\affiliation{Materials Science Division, Argonne National Laboratory, Argonne, Illinois 60439, USA}

\date{\today}
%\pacs{72.15.Jf, 73.63.-b, 85.80.Fi}

\begin{abstract}
We investigate the quantum heat exchange between a nanojunction and a many-body or electromagnetic environment far from equilibrium. It is shown that the two-temperature energy emission-absorption mechanism gives rise to a giant heat flow between the junction and the environment. We obtain analytical results for the heat flow in an idealized high impedance environment and perform numerical calculations for the general case of interacting electrons and discuss the giant freezing and heating effects in the junction under typical experimental conditions. 
\end{abstract}

\maketitle

Quantum dynamics of tunnel nanojunctions is governed by underlying relaxation mechanisms and nonequilibrium effects
since even small currents drive a nanojunction well out of equilibrium~\cite{Giazotto,Grabert_Devoret}.  At low temperatures the direct energy transfer to the phonon bath becomes inefficient and relaxation is dominated by the energy exchange between the tunnelling electrons
and electromagnetic environment and/or to many-body excitations in the electrodes (hereafter we will refer to both mechanisms
as to relaxation via the environment)~\cite{chtch}.  In this Letter we calculate the heat flow between the tunnelling electrons and the environment,
which control the junction dynamics, using a non-perturbative technique based on quantum kinetic equations taking into account
far from equilibrium effects.  We show, in particular, that a regime exists in which the interaction with the environment gives rise to
a giant cooling of the nanojunction.

\begin{figure}[hb]
  \includegraphics[width=0.75\columnwidth]{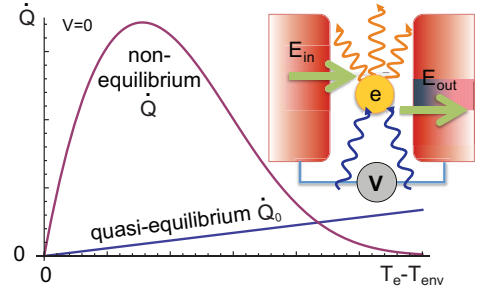}\\
  \caption{(color online) Illustration of the non-equilibrium heating effects in a nanojunction. The electrons traversing the junction absorb external photons (incident wavy lines) and emit them leading to heating of the contact. The plots show the {\it giant} heating effect as a function of the difference of electron and environment temperatures in the non-equilibrium situation compared to the quasi-equilibrium approximation at zero bias voltage. The full non-equilibrium analysis gives an at least one order of magnitude more pronounced heating effect than for the latter case: $\max ({\dot Q}/{\dot Q}_0) > 10$.}\label{fig.1}
\end{figure}

The energy exchange between tunnelling electrons and the environment is determined by the emission of  environment modes
with temperature equal to that of electrons, $T_{\mathrm e}$, and the absorption of environment excitations
carrying the temperature of the thermal bath, $T_{\mathrm{env}}$.
Moreover, not only temperatures, but also the distributions of emitted and absorbed environment modes may appear essentially different in the far from equilibrium regime $T_e\backsimeq V>T_{\rm env}\backsimeq T_{\rm leads}$,
where $V$ is the voltage across the junction.
Our main finding is that this two-temperature emission-absorption mechanism gives rise to a giant heat flow between the junction
and the environment (see inset in Fig.~\ref{fig.1}).

In the case of a resistive environment [electromagnetic fluctuations in a cold (hot) resistor shunting the tunnel junction]~\cite{Grabert_Devoret,Pekola,R_env}, the heat flow is
$\dot Q\backsimeq T_e R_T C\ln(R_T C/\tau_{\rm e})  k_B (T_{e}-T_{\rm env})/R_T C$, with comparable temperatures $|T_{e}-T_{\rm env}|\ll T_{e}+T_{\rm env}$. Here $R_T$ and $C$ are the ohmic resistance~\cite{note} and
capacitance of the junction, respectively, and
$\tau_{\rm e}$ is the electron energy relaxation time, $\tau_{\rm e}\ll1/T_e\ll R_T C$.
This result well exceeds the flow in the
quasi-equilibrium approximation $\dot Q_0\backsimeq k_B (T_{e}-T_{\rm env})/R_T C$~\cite{Pekola}, where the emitted and absorbed modes have the same temperature. The large factor $T_e R_T C\ln(R_T C/\tau_{\rm e})\gg 1$ by which the two results differ, reflects the elevated effective number of environment excitations emitted by charges tunnelling through the nanojunction in the out of equilibrium regime, see  Fig.~\ref{fig.1}.

\paragraph{Model}
The rate of the heat flow between the  tunnel junction and the environment is given by~\cite{note2}:
\begin{gather}\label{eq:Q}
    \dot{Q}=\int_{0}^\infty\,\varepsilon\left\{n_\varepsilon P(\varepsilon)-[1+n_\varepsilon] P(-\varepsilon)\right\} p(\varepsilon)d\varepsilon,
\end{gather}
where $P(\pm\varepsilon)$ is the probability density for the tunneling charge-carrier to lose [gain] the energy $\varepsilon$ to [from] the environment.
\begin{figure}[ht]
  \includegraphics[width=0.8\columnwidth]{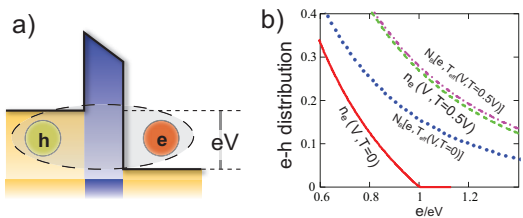}\\
  \caption{(color online) a) Illustration of electron-hole pair generation in the tunnel junction, resulting in the distribution function $n_\varepsilon$ [Eq.~(\ref{eq:Q})] of these pairs (environment). b) Comparison of the distribution functions for $T=0$ and $T=V/2$.}\label{fig.eh}
\end{figure}
The distribution function $n_\varepsilon$ in Eq.~(\ref{eq:Q}) can be interpreted as the distribution function of electron-hole pairs that appear at the junction interface just after the tunneling process: the hole in the source lead and the electron in the drain, Fig.~\ref{fig.eh}a). If the distribution functions at the electrodes are Fermi functions with equal temperatures $T_e$, then $ n_\varepsilon = \{(\varepsilon-V)N_B(\varepsilon-V,T_e)+(\varepsilon+V)N_B(\varepsilon+V,T_e)\}/2\varepsilon$, with $N_B(\varepsilon \pm V, T_e)$ being the equilibrium Bose distribution function.
For this case the effective temperature of the nanojunction is defined as
\begin{gather}
\label{Teff}
    T_{\rm eff} (V, T_e) = \lim_{\varepsilon\to 0} n_\varepsilon(V, T_e) = \frac{V}{2} \coth\frac V{2T_e}.
\end{gather}
At low applied voltages, $V \ll T_e$, the effective temperature of the junction $T_{\rm eff}$ coincides with the temperature of the leads, $T_{\rm eff} \approx T_e$. In the opposite case of high voltages, $V \gg T_e$, we obtain $T_{\rm eff} \approx V/2$.
The function $p(\varepsilon)$ in Eq.~(\ref{eq:Q}) is the weight function for a junction between two normal metals, Fig.~\ref{fig.1}, and can be calculated for any choice of the electron distribution function in the leads, resulting in $p(\varepsilon) = 4\varepsilon/R_T$.

\paragraph{Heat flow}
To calculate $\dot{Q}$ one has to specify the
probability density, which can be written in a form $P(\varepsilon)=\int_{-\infty}^\infty dt \exp[J(t) + i\varepsilon t]$,
where the function $\exp[J(t)]$ reflects the fact that tunneling electrons acquire random phases due to interaction with the
Bosonic environment, represented by a set of oscillators with non-equilibrium distribution of modes, $N_{\omega}$.
The quasi-equilibrium situation where the distribution functions of the environment modes
are Bose distributions parametrised by equilibrium temperatures was discussed in Ref.~\cite{Grabert_Devoret}.
In general far from equilibrium, $J(t)$ is~\cite{chtch}:
\begin{equation}\label{eq:J}
J(t)=2\int\limits_{\tau_e^{-1}}^\infty \frac{d\omega}{\omega}\rho(\omega)
\left[N_\omega^{(\rm in)} e^{i\omega t}+(1+N_\omega^{(\rm out)})e^{-i\omega t}-B_\omega\right]\,.
\end{equation}
The mode distribution, $N_{\omega}$, is defined by a kinetic equation with scattering integral
describing the energy exchange between environment modes and tunnelling electrons.
The terms proportional to $N_\omega^{(\rm in)}$ and $1+N_\omega^{(\rm out)}$ correspond to the
absorbed and emitted environment excitations, respectively.
The combination $B_\omega = 1 + N_\omega^{(\rm out)} + N_\omega^{(\rm in)}$ is the kernel of the time-independent contribution
to $J(t)$ describing the elastic
interaction of the tunnelling electrons with the environment modes.
In equilibrium $N_\omega$ reduces to the Bose-function and the functional $P(\omega)$ recovers the result of
Ref.~\cite{Grabert_Devoret}. In Eq.~(\ref{eq:J}), the energy relaxation time $\tau_e$ determines the low energy cut-off,
since the electrons start to equilibrate on larger time scales, i.e. the non-equilibrium description does not hold any more.
The spectral function $\rho(\omega)$ is the probability of the electron--environment
interaction and characterizes the particular system under consideration.

To estimate the magnitude of the heat flow $\dot{Q}$ we first expand the distribution function
$P(\varepsilon)$ in Eq.~(\ref{eq:Q}), in the first order in $\rho(\varepsilon)$:
\begin{equation}\label{perturb1}
    \dot{Q}^{(1)}=\frac{4}{R_{\scriptscriptstyle {\mathrm T}}} \int\limits_{\tau_e^{-1}}^\infty \frac{d\varepsilon}{2\pi} \varepsilon\rho(\varepsilon)
\left\{ n_{\varepsilon}( 1 + N_{\varepsilon}^{\rm(out)}) - (1 + n_\varepsilon)N_{\varepsilon}^{\rm(in)}\right\}.
\end{equation}
The expression in Eq\,(\ref{perturb1}) becomes zero if  $n_\varepsilon = N_\varepsilon^{\rm(in)} = N_\varepsilon^{\rm(out)}$.
If the distribution functions are not equal to each other, we can expand $\dot{Q}^{(1)}$ with respect to their difference.
We consider the case where the voltage bias at the nanojunction is zero but the temperatures of electrons at the
leads and those that comprise the environment are slightly different, $T_{\mathrm{e}}=T+\delta T/2$ and
$T_{\mathrm env}=T-\delta T/2$.
Thus,
$n_\varepsilon = n_\varepsilon(T+\delta T/2)$, $N_\varepsilon^{\rm(in)} = n_\varepsilon(T-\delta T/2)$, $N_\varepsilon^{\rm(out)} =
n_\varepsilon(T+\delta T/2)$, where $n_\varepsilon$ is the Bose distribution function.
Using Eq.~(\ref{perturb1}) in the first order in small parameter $\delta T/T \ll 1$ we find
\begin{gather}\label{eq:Q1dT}
    \dot{Q}^{(1)}_\theta\approx \delta T\frac4{R_{\scriptscriptstyle {\mathrm T}}} \int\limits_{\tau_e^{-1}}^\infty \frac{d\varepsilon}{2\pi} \varepsilon\rho(\varepsilon)
    n_\varepsilon'(T)(1 + \theta n_\varepsilon(T))\, ,
\end{gather}
where $n_\varepsilon'(T) = d n_\varepsilon(T)/d\varepsilon$. The index $\theta$ is $0$ for the quasi-equilibrium situation when the temperatures of emitted and absorbed environment excitations are equal and $1$ for the non-equilibrium case (the index $1$ is skipped throughout this Letter). Since $n_\varepsilon(T)$ in Eq.~(\ref{eq:Q1dT}) is always positive, the following inequality is valid $|\dot{Q}^{(1)}_0|<|\dot{Q}^{(1)}|$, where $\dot{Q}^{(1)}_{0}$ and $\dot{Q}^{(1)}$ refer to the heat flux in quasi-equilibrium and in non-equilibrium cases, respectively. The interaction function $\rho(\varepsilon)$ in Eq.~(\ref{eq:Q1dT}) quickly decays at frequencies larger than some characteristic frequency $\omega_{\rm max}$. For temperatures $T>\omega_{\rm max}$ (quantum regime) we can approximate $n_\varepsilon (T) \approx T/\varepsilon \gg 1$ and find
\begin{gather}%\label{}
\frac{|\dot{Q}^{(1)}|}{|\dot{Q}^{(1)}_{0}|}\approx\frac{\int_{\tau_e^{-1}}^\infty \frac{T\rho (\omega)d\omega}\omega}{\int_{\tau_e^{-1}}^\infty \rho (\omega)d\omega}\approx \frac{T}{\omega_{\rm max}}\ln (\omega_{\rm max}\tau_e)\gg 1.
\end{gather}
Remarkably, in higher orders with respect to $\rho(\varepsilon)$ the non-equilibrium heat flow
 $\dot{Q}$ differs from the equilibrium flow $\dot{Q}_{0}$ by the same factor (see supplementary material). This result holds even for a finite electric current flowing through the junction. However, in this case we need to replace the temperature $T$ by the effective temperature $T_{\rm eff}$,
see Eq.~(\ref{Teff}), of the tunnelling electrons. Thus, the heat flow between the junction and the environment appears
much larger than what the quasi-equilibrium estimates predict.

\paragraph{Ohmic approximation}
\begin{figure}[t]
\includegraphics[width=0.9\columnwidth]{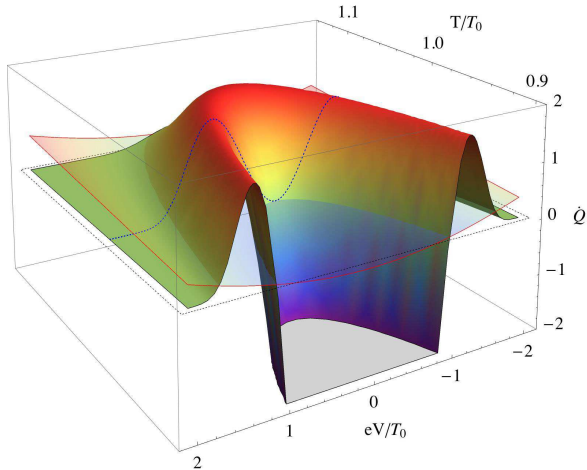}\\
  \caption{(color online) Typical heat exchange $\dot{Q}$ in Eq.~(\ref{eq:Q}) of the Ohmic environment with the tunnel junction between two normal leads. $\dot Q(T_{\rm eff},T,V)$ vs $T/T_0$ and voltages $eV/T_0$ ($T_0=30\omega_{\rm max}$.). We used $\omega_{\rm max}/\omega_{\rm min}=100$, $T_{\rm env}/T_0=1$, $\rho(0)=10$. $\dot Q$ is measured in units of $10^3\omega_{\rm max}^2/(e^2R_T)$. }\label{fig.Q3Dohm}
\end{figure}
We now turn to the simplest case, an environment with a
very high impedance as compared to the quantum resistance, $R_{\mathrm{Q}}$.
In this limit tunnelling electrons easily excite the environment modes.
The spectral density $\rho(\omega)$ of these modes is sharply peaked at the zero frequency, $\omega = 0$.
For the correlation function $J(t)$ in Eq.~(\ref{eq:J}) the concentration of the environment modes at low frequencies
implies that the expansion of $J(t)$ over $t$ up to the second order yields $J(t)\approx - i a t - (b/2) t^2$,
where the coefficients $a$ and $b$ are defined as
$a = \int_{\tau_e^{-1}}^{\infty}( 1 + N_\omega^{\rm (out)} - N_{\omega}^{\rm (in)})\rho(\omega) d\omega$ and
$b = \int_{\tau_e^{-1}}^{\infty}\omega \rho(\omega) B_\omega d\omega$.
Using this expansion for $J(t)$ we obtain the following result for the density function $P(\omega)$
\begin{gather}
\label{P1}
P(\varepsilon) = (1/\sqrt{2\pi b}) \exp \left[- (\varepsilon - a)^2 /2b\right].
\end{gather}
Here the expansion parameter $a$ can be estimated as follows
$a = a_0\left( 1 +\frac{(T_{\rm e} - T_{\rm env}) \ln(\omega_{\rm max}\tau_e)}{\pi \omega_{\rm max}}\right)$,
where  $a_0=2\int \rho d\omega \approx 2\rho(0)\omega_{\rm max} \approx 2E_c$ with $E_c$ being the charging energy
of the tunnel junction, $T_{\rm e}$ is the electron temperature in the junction, $T_{\rm env}$ is the temperature of environmental modes, $\omega_{\rm max} \approx 1/(R_T C)$. Similar for coefficient $b$ in Eq.~(\ref{P1}) we obtain $b \approx a_0 (T_{\rm e} + T_{\rm env})$.

Substituting the density  $P(\omega)$, Eq.~(\ref{P1}), into  the heat
flux $\dot{Q}$, Eq.~(\ref{eq:Q}), we obtain our first main result for the typical heat exchange of the Ohmic environment with the tunnel junction between two normal leads. The full temperature and voltage dependence is shown in Fig.~\ref{fig.Q3Dohm}.

\paragraph{Dynamic Coulomb interaction}
\begin{figure}[b]
 \includegraphics[width=0.7\columnwidth]{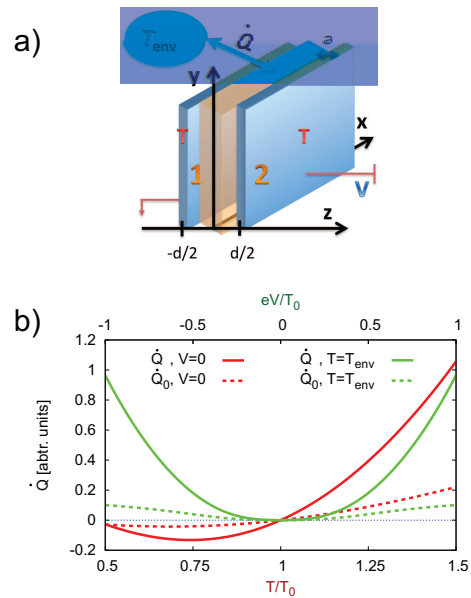}\\
  \caption{(color online)
 {\bf  a)}: Schematic presentation of the system: single contact junction, with contacts consisting of two thin plates, which are distance $d$ apart. Their thickness $a$ is much less than the extension in $x$ and $y$ directions, such that they can be treated as 2D contacts. The temperature of the contacts $T$ is kept constant, while the environment temperature $T_{\mathrm{env}}$ can be different, which results in heat production or removal in the junction.
 {\bf b)}: Heating of a tunnel junction taking into account dynamic Coulomb interactions for the zero bias case ($V=0$) [red lines, lower x-axis] and the voltage dependence for $T=T_{\rm eff}$ [green lines, upper x-axis]. The solid cures represent the quasi-equilibrium curves and  the dashed assuming an equilibrium distribution for $N_\omega$. ($T_0=0.1E_{\rm th}$, see text also)}\label{fig.DCI}
\end{figure}
Next we discuss the more realistic situation where the tunneling
junction is connected to two disordered conductors (leads). Following Ref.~\cite{Grabert_PRL}, one
can find the spectral probability function $\rho(\omega)$ in Eq.~(\ref{eq:J}) corresponding to the electron--environment
interaction
\begin{equation}
\label{rho2}
\rho_{\mathrm{ij}}(\omega)
=\frac{\omega}{2\pi}\Imm\sum_{\mathbf q}\frac{\left(\frac{2\pi}{L}\right)^2(2\delta_{\rm ij}-1)\tilde U_{\rm ij}({\mathbf q},\omega)}
{(\D_{\rm i} q^2-i\omega)(\D_{\mathrm j} q^2-i\omega)}\,,
\end{equation}
where $i,j=1,2$ are the lead indices, $\D_{1(2)}$ are diffusion coefficients within respective electrodes,
and $\tilde U_{\rm ij}({\mathbf q},\omega)$ are the dynamically screened Coulomb interactions within (across) the electrodes.
The form of spectral probability $\rho(\omega)$ [$\rho(\omega)=2\rho_{12}+\rho_{11}+\rho_{22}$] depends on the structure of the environmental excitations spectrum and, thus, on the external bias.

The system under consideration is shown in Fig.~\ref{fig.DCI}a): two contacts are separated by distance
$d$ and their thickness is $a$. The external bias is $V$ and the contacts are kept at
temperature $T$ and the environment at temperature $T_{\rm env}$. Two situations are possible:
i) for zero bias, $V=0$ we have $T_{\rm out}=T_e=T_{\rm eff}=T$ and $T_{\rm in}=T_{\rm env}$. ii) for $V\neq 0$ the
effective temperature $T_{\rm eff}$ depends on $V$ as shown in Eq.~(\ref{Teff}).

The screened Coulomb interaction in Eq.~(\ref{rho2}) in Fourier space has the form $\underline{\tilde U}(\mathbf{q},\omega) = \{[\underline{U}^{(0)}(\mathbf{q},\omega)]^{-1}+{\underline{\mathcal{P}}}(\mathbf{q},\omega)\}^{-1}$,
where $\underline{U}^{(0)}(\mathbf{q},\omega)=u(q)\underline{I}+v(q)\underline{\sigma}_x$ is the bare Coulomb interaction and $\underline{\mathcal{P}}(\mathbf{q},\omega)$ the
polarization matrix respectively with $\mathcal{P}_{\rm ij}=\nu_i \D_i q^2(\D_i q^2-\imath\omega)^{-1}\delta_{\rm ij}$.
$\nu_i$ is the electron density of states at the Fermi surface in lead $i$.

Below we concentrate on quasi 2D infinite leads.
For this geometry with $a\ll L$, where $L$ is the characteristic lead size in the $x$ and $y$ directions,
the bare Coulomb interaction has the form
\begin{equation}
U_{ij}^{(0)}(\mathbf{r}_i-\mathbf{r}_j)=e^2\int dz_i\, dz_j\, \frac{\delta(z_i-z^{(0)}_i)\delta(z_j-z^{(0)}_j)}{|\mathbf{r}_i-\mathbf{r}_j|}\,,
\end{equation}
with $z^{(0)}_i=(1/2-\delta_{i1})d$, leading to $u(q)=2\pi e^2/q$ and $v(q)=2\pi e^2 e^{-qd}/q$.

In the following, we consider the case of identical leads with same diffusion coefficients $\D_1 = \D_2 \equiv \D$ and densities of states, $\nu_1 = \nu_2 \equiv \nu$.
The dimensionless matrix elements $\tilde U_{ij}$ of the dynamically screened Coulomb interaction (in units of $e^2 d$) are then given by
\begin{equation}
\tilde U_{\rm ii} = \frac{4 \pi}{\tilde{q}}  \frac{\chi(\tilde{q})}{\chi^2(\tilde{q})- \coth^{-2}(\tilde{q})}\,,\,\,  \tilde U_{\rm i\neq j}=\frac{\tilde U_{ii}}{\chi(\tilde{q})\coth(\tilde{q})}\label{tildeU}
\end{equation}
where $\tilde{q}=dq$ and $\tilde\omega\equiv \omega (d^2/\D)$ with the dimensionless
function $\chi(\tilde{q}) \equiv 1+\coth(\tilde{q})+\frac{4\pi e^2 d\nu  x}{ \tilde{q}^2 - i \tilde\omega}$.
Using these expressions, we can write Eq.~(\ref{rho2}) as
\begin{equation}
\label{rho3}
\rho(\tilde\omega) = \frac{2 e^2 d}{\D} \tilde\omega \Imm  \int\limits_0^\infty \tilde{q} d \tilde{q}  \frac{\tilde U_{11} \left[1-\left(\chi(\tilde{q})\coth(\tilde{q})\right)^{-1}\right]}{(\tilde{q}^2 - i\tilde\omega)^2}\,.
\end{equation}

Substituting Eq.~(\ref{rho3}) into Eqs.~(\ref{eq:J}) we can calculate the heat flux $\dot{Q}$ in Eq.~(\ref{eq:Q}) between environment and nanojunction with dynamic Coulomb interaction.
The typical energy scale is given by the Thouless energy for the junction of distance $d$, $E_{\rm th}={\cal D}/d^2$ which we use to rewrite all expressions in dimensionless units.
For a typical temperature $T_0=0.1E_{\rm th}\approx 10K$, the temperature and voltage dependence is numerically calculated and shown in Fig.~\ref{fig.DCI}b). Again, the non-equilibrium heat flow $\dot Q$ is up to an order of magnitude larger and the quasi-equilibrium approximation $\dot Q_0$. We remark, that in this case the function $\rho(\omega)$ introduces a natural cut-off for $J(t)$, Eq.~(\ref{eq:J}), which behaves as $\sim -|t|$ for large $t$. % in contrast to the Ohmic approximation.

\paragraph{Discussion}
Above we assumed that hot electrons interact with acoustic phonons (acoustic
environment modes). This assumption holds if the environment temperature $T_{\rm env}$ is lower
than the Debye temperature
$\Theta_D$, which is of the order of optical phonon energies. In this temperature range the electron
interaction with the environment is quasi-elastic because the change of the electron energy, which is equal
to the energy of the emitted or absorbed phonons is much smaller than the electron energy. Due to
the small inelasticity of the acoustic phonon scattering, the deviation of the electron distribution
function in the momentum space from the isotropic one is small even when the electrons become hot.

We also assume that the density of hot electrons is high enough so that the electron-electron
scattering time $\tau_{e-e}$ is smaller than the time of energy relaxation $\tau_{\rm env}$ (this time
is large because of quasi-elastic nature of interaction between the electrons and environment). In this
case the electron distribution function is close to an equilibrium one with an electron temperature $T_e$,
which is in high voltage limit is higher than the environment temperature $T_{\rm env}$. At very high
applied voltages the electron energies become comparable with the energies of optical phonons (optical
environmental modes) and the approximation of small inelastic scattering does not hold.

In summary, we discussed the influence of far from equilibrium heating effects on properties of nanojunctions.
Based on a quantum-kinetic approach we calculated the non-linear heat flux between environment and junction. We showed
that the resulting freezing or heating effect far from equilibrium are by orders of magnitude larger than estimates based on
quasi-equilibrium environment theory. We obtained analytical results for the heat flow in an idealized high-impedance environment and demonstrated, numerically, that these results hold for the more general case of an environment with Coulomb interaction.  
We showed that the environment can be a very effective freezing agent if the effective temperature well exceeds the high frequency cut-off $\hbar\omega_{\rm max}$.

One can expect that our results, in particular the giant freezing effect, will be important for the electronic transport in junction arrays~\cite{grains}, which will be subject of a forthcoming work.

This work was supported by the U.S. Department of Energy Office of Science under the Contract No. DE-AC02-06CH11357.
I.~B. was supported by an award from Research Corporation for Science Advancement and the Materials Theory Institute at ANL.


\begin{thebibliography}{99}
\bibitem{note}
Since the current-voltage characteristic $I(V)$ becomes linear only at high voltages,
$R_{\mathrm{T}}$ is defined at large $V$.
\end{thebibliography}
\end{document}